\documentclass[12pt]{article}

 \normalsize

\usepackage{scicite}
\usepackage{color}
\usepackage{times}
\usepackage{amsmath}
\usepackage{amssymb} 
\usepackage{graphicx}
\usepackage{xcolor}
\usepackage{verbatim}
\usepackage{bm}
\usepackage{textcomp}

\newcommand{\etal}{\textit{et al}.}

\title{Examining mobility data justice during 2017 Hurricane Harvey}

\author
{Hengfang Deng,$^{1}$ Qi Wang$^{1, 2\ast}$ \\
\\
\normalsize{$^{1}$Civil and Environmental Engineering, Northeastern University, Boston 02115, USA}\\
\normalsize{$^{2}$School of Public Policy and Urban Affairs, Northeastern University, Boston 02115, USA}\\
\normalsize{$^\ast$ q.wang@northeastern.edu}
}

\date{}

\begin{document}

\maketitle

\section*{ABSTRACT}
Natural disasters can significantly disrupt human mobility in urban areas. Studies have attempted to understand and quantify such disruptions using crowdsourced mobility data sets. However, limited research has studied the justice issues of mobility data in the context of natural disasters. The lack of research leaves us without an empirical foundation to quantify and control the possible biases in the data. This study, using 2017 Hurricane Harvey as a case study, explores three aspects of mobility data that could potentially cause injustice: representativeness, quality, and precision. We find representativeness being a major factor contributing to mobility data injustice. There is a persistent disparity of representativeness across neighborhoods of different socioeconomic characteristics before, during, and after the hurricane's landfall. Additionally, we observed significant drops of data precision during the hurricane, adding uncertainty to locate people and understand their movements during extreme weather events. The findings highlight the necessity in understanding and controlling the possible bias of mobility data as well as developing practical tools through data justice lenses in collecting and analyzing data during disasters.\par

\textbf{Keywords:} Human mobility; Data justice; Hurricane Harvey; Neighborhood; Disaster informatics
 
\section{INTRODUCTION}

Mobility, i.e., the ability and capacity to move and travel, is a basic human behavior \cite{adey2010aerial,cresswell2011geographies}. Natural disasters can significantly disrupt human mobility\cite{wang_patterns_2016,wang_quantifying_2014} and cause displacements for evacuation and sheltering purposes \cite{song2013,Yabe2020,Kashif2013}. From 2007 to 2017, natural disasters have caused over 2.4 billion people being displaced and about 31 million people homeless \cite{ritchie_natural_2014}. When Hurricane Irma was approaching, about 30\% of Florida residents evacuated. A survey conducted after the strike of Hurricane Florence showed that 22\% of Americans have ever had to evacuate their homes due to the threat of a natural disaster\cite{Ballard2018}. The intensifying climate change and high-frequency of extreme weather are likely to increase the severity of human mobility disruptions. \par 

Crowdsourced data collected from social media and cell phones has been used to study human mobility (see ~\cite{barbosa_human_2018} for a comprehensive review). These data sets provide unprecedented quantity, resolution, and sample sizes which significantly surpass traditional data sets based on surveys, interviews, or travel diaries. Also, findings based on these data have shed new insights on human movements under the influence of natural disasters~\cite{wang_patterns_2016,wang_quantifying_2014,song_prediction_2014}. Beyond, mobility disruptions have been linked to social vulnerability of urban neighborhoods and demonstrated promising results in emergency response and evacuation behaviors~\cite{yabe_effects_2020,yabe_cross-comparative_2019,yabe_understanding_2020,kumar_utilizing_2018,martin_leveraging_2017}.\par

However, there is a growing concern about the justice of crowedsourced data sets \cite{taylor2017data}, which can be caused by technology inequalities deeply rooted in the socio-economic system. Recent studies find a substantial “digital divide” between racial and ethnic groups in the US in smartphone uses \cite{warschauer2004,monroe2004}. In mobility-related research, the divide can cause the injustice in representativeness: a significant scarcity of individuals from underrepresented and vulnerable communities. Consequentially, engineering solutions based on these data can be biased, causing an underestimate or even negligence of these populations' needs. The injustice could worsen during the occurrences of disasters \cite{Femke2020Humanitarian}, undermining the effectiveness of models and algorithms designed for disaster responses and relief. Therefore, an accurate understanding and prediction of human mobility, which depends on unbiased and accountable access to representative, rich, and high-quality data sets, can be a matter of life and death during natural disasters. \par

Although a few studies have examined the changes of crowdsourced data sets from social media, especially Twitter, during natural disasters \cite{fan2020spatial,Zou2019Social,samuels2020silence,kumar_enhancing_2020,hao2020leveraging}, limited research has studied the justice issues of mobility data in the context of natural disasters. The lack of research leaves us without an empirical foundation to quantify and control the possible biases in the data and prohibits the design of effective and inclusive algorithms or \textit{fairness artificial intelligence} (FAI). In this study, we take a first step to quantify mobility data justice. The study is a building block to accurately predict human mobility during hurricanes and advance our ability to provide individualized risk assessment and disaster support for urban dwellers, especially those from unrepresentative and disadvantaged neighborhoods. \par 

\section{BACKGROUND}
\subsection{Human mobility and its applications in social studies}

A considerable amount of research has examined and improved our understanding of human mobility patterns. Human mobility has been proved to possess some fundamental and universal patterns. These patterns include high uniformity \cite{gonzalez2008understanding,jurdak2015understanding,brockmann2006scaling}, ultraslow diffusion \cite{song2010modelling,song2010limits,Toole2015Coupling,cuttone2018}, periodicity\cite{cho2011friendship,LIANG20122135}, high predictability\cite{schneider_unravelling_2013,wang2017predicting}, and motif composition\cite{schneider_unravelling_2013,Schneider2013,candia2008uncovering}. \par

Recently, researchers started to link human mobility with social phenomena. Kwan \cite{Kwan2013Beyond} pointed out that mobility is an essential element of people's spatiotemporal experiences and it should be part of the integrated analysis to examine people's everyday experiences. Indeed, human mobility studies in multiple global cities have found socioeconomic characteristics are strongly associated with urban dwellers' navigation in the metropolitan areas \cite{bajardi2015unveiling,wang2018urban,long2015profiling}. Ruktanonchai \etal \cite{ruktanonchai2018using} studied the link between human mobility and socioeconomic development. They found that the diversity of mobility patterns are correlated with the external socioeconomic indicators. Šćepanović \etal \cite{scepanovic2015mobile} showed that mobility patterns have a high correlation with many of the socioeconomic factors, revealing a diversity of attributes that can be inferred using mobile phone call data.\par

A few studies have examined how human mobility is impacted by social segregation. Palmer \etal \cite{palmer2013new} developed a pilot study called the Human Mobility Project to collect GPS data from smartphones and conducted surveys on activity spaces, social segregation, and subject well-beings of the participants. It is one of the pioneering works that aim to move spatial measures beyond the residential census unit. The study linked human mobility and demographic research, although the small sample sizes, admitted by the authors themselves, were likely to introduce bias and incompleteness. Amini \etal \cite{amini2014impact} studied mobility patterns from Ivory Coast and Portugal and reported significant differences in the two regions. They argued that cultural and linguistic diversity in developing regions could constrain mobility options from vulnerable populations. Wang \etal \cite{wang2018urban} studied human mobility and urban isolation across 50 cities in the United States. The study found that while residents from different types of neighborhoods are likely to travel similar distances and to similar numbers of neighborhoods, people from disadvantaged neighborhoods are less likely to travel to non-poor white ones. A follow-up study from this team \cite{Phillips2019social} proposed developing network-based measures of “structural connectedness” based on the everyday travel of people across neighborhoods collected from mobility datasets. They demonstrated that the connectedness indices could capture not only the equity of mobility among neighborhoods but also the concentration of mobility patterns within a city \cite{Phillips2019social}. Lathia \cite{lathia2012hidden} used public transport fare as a proxy to study London urban flow and its correlation with urban dwellers' well-being. They found that deprived areas tend to preferentially attract people living in other deprived areas, suggesting a segregation effect. \par

Some studies, however, have found limited support that socioeconomic status impacts human mobility patterns. For example, Xu \etal \cite{xu2018human} studied human mobility in both Singapore and Boston using large-scale mobile phone data sets. They examined six mobility indicators that are associated with socioeconomic status, namely radius of gyration, number of activity locations, activity entropy, travel diversity, k-radius of gyration, and unicity. They found that phone users across different socioeconomic classes (albeit with a focus on wealth) exhibit very similar characteristics for both cities. \par

Human mobility has also been linked to the well-being of different populations. Bosetti \etal \cite{bosetti2019reducing} studied the mobility patterns of Turkish and Syrian refugees in Turkey and found that somewhat counter-intuitive yet concrete evidence that social segregation could boost potential outbreaks of measles. Therefore, policies that encourage integration are needed to reduce the transmission of diseases. A report from New York Times \cite{valentino-devries2020location} during the COVID-19 pandemic also found that residents from the most impoverished neighborhoods took at least three more days to reduce their mobility (i.e., practice social distances) than the ones from the wealthiest neighborhoods. Decuyper \etal \cite{decuyper2014estimating} showed that the proxies derived from mobile phone data could provide valuable up-to-date operational information on food security throughout low and middle-income countries. \par

\subsection{Human mobility during natural disasters}

There has also been a large body of research utilizing human mobility data to investigate disaster-related response and preparedness activities primarily in three perspectives: individual and collective perturbation and resilience patterns, predictive modeling of human emergency behavior, and its association between demographic, socioeconomic, and other factors. \par

Previously, several studies \cite{bagrow2011collective,gray2012natural,lu2012predictability,wang2014quantifying} have found that large-scale natural disasters such as hurricanes and earthquakes could disrupt human mobility patterns. Wang and Taylor \cite{wang2016patterns} compared mobility changes during five types of natural disasters using Twitter data and discovered that the power-law could capture even the disrupted human mobility patterns. Furthermore, it found that the pre-disaster and post-disaster mobility patterns correlated. Wang \etal \cite{wang2017aggregated} investigated the human mobility changes during the 2015 severe winter storms in the Northeast region in the U.S. The analysis of over 2.6 million geotagged Tweets demonstrated that both the distances of spatial displacement and radii of gyration of individuals’ mobility were disrupted significantly. Also, travel patterns pre-disaster became irregular during the extreme weather event. Martin \etal \cite{martin2017leveraging} examined the spatiotemporal changes in human behavior using Twitter data after Hurricane Matthew and confirmed the effectiveness of using such crowd-sourced data for disaster awareness and evacuation behavior analysis. Finally, Ahmouda \etal \cite{ahmouda2019using} demonstrated the resilience of mobility across the population using Twitter data following the 2016 Hurricane Matthew and 2017 Hurricane Harvey. The authors found that displacements became shorter and areas of activity became smaller during hurricanes, although power-law models could still approximate the distribution of displacements. In addition to understanding the perturbation effect on the individual level, several studies also investigated the urban community network change following the disaster. Sadri \etal \cite{sadri2018analysis} presented a framework to build social networks during the 2012 Hurricane Sandy with Twitter data and showed that the user degrees follow the power-law distribution. Another study \cite{sadri2020exploring} found that social interaction networks increased assortativeness for significant subgraphs. \par

Several studies explore the predictive modeling of emergency behavior under extreme disasters. Aschenbruck \etal \cite{aschenbruck2004human} proposed a gravitational simulation model of mobility in disaster-affected areas. The gravity-based approach demonstrates higher accuracy than those derived from Gauss-Markov and random walk mobility models. Song \etal \cite{song2013} developed probabilistic inference models to capture collective mobility patterns during a disaster to better understand evacuation behaviors under disasters. The models achieved an overall of 80\% accuracy. Follow-up studies from Song \etal \cite{song_prediction_2014, song2015simulator} aggregated a large number of individual trajectories after the Great East Japan Earthquake and the Fukushima Daiichi nuclear disaster and built population mobility graphs through collaborative learning. Markov Decision Processes (MDPs) were then used to train and predict people’s new locations (e.g., shelters, etc.) after the earthquake. The matching rate for post-disaster movements was approximately 63.18\%, and their models outperformed other traditional methods such as HMMs(Hidden Markov Models). More recently, in further development by the same group \cite{song2017deepmob}, deep learning models were introduced, which improved the matching rate to 77.58\%, demonstrating the possibility of increasing the accuracy of predicting human mobility after perturbations caused by natural disasters. \par
A few studies also modeled the mobility changes in transportation networks. Nelson \etal \cite{nelson2007event} proposed a gravity-based role-based mobility model to describe network recovery from natural disasters. Uddin \etal \cite{uddin2009post} developed a Delay-Tolerant Networking (DTN) embedded mobility model capturing disasters' impact on the urban transportation network. Aschenbruck \etal \cite{aschenbruck2009modeling} proposed an area-based approach for modeling objects' movements such as vehicles in the affected regions.  Finally, Nadi \etal \cite{nadi2017adaptive} developed a multi-agent assessment and response system (MARS) simulating the interaction between human and intelligent agents, which embeds the Markov decision process in an evacuation demand and response. \par
 
Other studies have also analyzed the human response to disaster and recovery process on the neighborhood level and identified the collective mobility pattern’s association with community-level socioeconomic indicators and other regional factors. Yabe \etal \cite{yabe2019mobile} analyzed large scale mobile phone data collected from Puerto Rico and revealed the importance of inter-city social connectivity on evacuation decisions and disaster recovery after hurricanes and earthquakes. Yabe \etal also \cite{yabe_understanding_2020} found that population recovery patterns follow a universal negative exponential function, and the rates of evacuation and recovery could be associated with sociodemographic variables such as population size, median income, infrastructure damage level, and proximity to other cities. Metaxa-Kakavouli \etal \cite{metaxa2018social} studied the correlation between evacuation activities and close social ties using mobility data from over 1.5 million social media users following multiple hurricanes. Their results highlighted social ties play an important role in evacuations. Recently, Yabe \etal \cite{yabe_effects_2020} combined mobility data from over 1.7 million smartphones with income information from the census to understand the effects of income inequality on human emergency behavior during Hurricane Irma. They found that residents from wealthy communities were more likely to evacuate from the flooded areas and relocated to the regions that were not exposed to infrastructure damage risks. Hong \etal \cite{hong2020modeling} used large-scale Twitter data during 2017 Hurricane Irma to quantify evacuation flows at multiple geographical scales. In Florida, the most affected state, evacuation flows are well predicted by distances between geographical units as well as socioeconomic similarities. Collectively, these studies highlight the importance and yet the heterogeneous impacts of the demographic and socioeconomic factors on human mobility patterns under the influence of disasters. \par

\subsection{Mobility data sets and their justice}
Mobility data primarily from four different resources have been used to arrive at these new findings. They are: (1) GPS locations collected by GPS loggers or GPS-phones which have created some open and standard data sets for relatively small populations (i.e., 100-200); (2) CDR (call detailed records) collected from cell phone users when they text or make phone calls; (3) social media data collected from these platforms if they have a geolocating function; and (4) LBS (location-based services) data from the service providers who embed their geolocating functions in many smartphone apps. \par

Despite the variety, these data sets suffer from the same issue: unknown justice. Such an issue can come from complex causes and vary by different data sets. For example, Twitter, a popular social media platform, tend to attract younger users who are more likely to be males and less representative of minorities. Ruktanonchai \etal \cite{ruktanonchai2018using} used Google Location History to study human mobility but acknowledged that more data could be from middle and lower-income classes, which could potentially introduce bias. Beyond representativeness issues, justice can be undermined by uneven data quantity (e.g., no. of records) and quality (e.g., resolution and accuracy). These two factors become a more significant concern during natural disasters. Vulnerable populations from disadvantaged and underrepresented neighborhoods are more likely to suffer damages (e.g., flooding, power outages, etc.), making them more likely to send fewer data points. Moreover, the quality of the data is more likely to deteriorate by the disruptions. In this case, mobility-related measures and solutions are more likely to overlook their pressing needs when facing external shocks. \par

\section{METHODOLOGY}

\subsection{Data Sets}

\begin{figure}[htbp]
    \centering
    \includegraphics[page=1,width=0.6\textwidth]{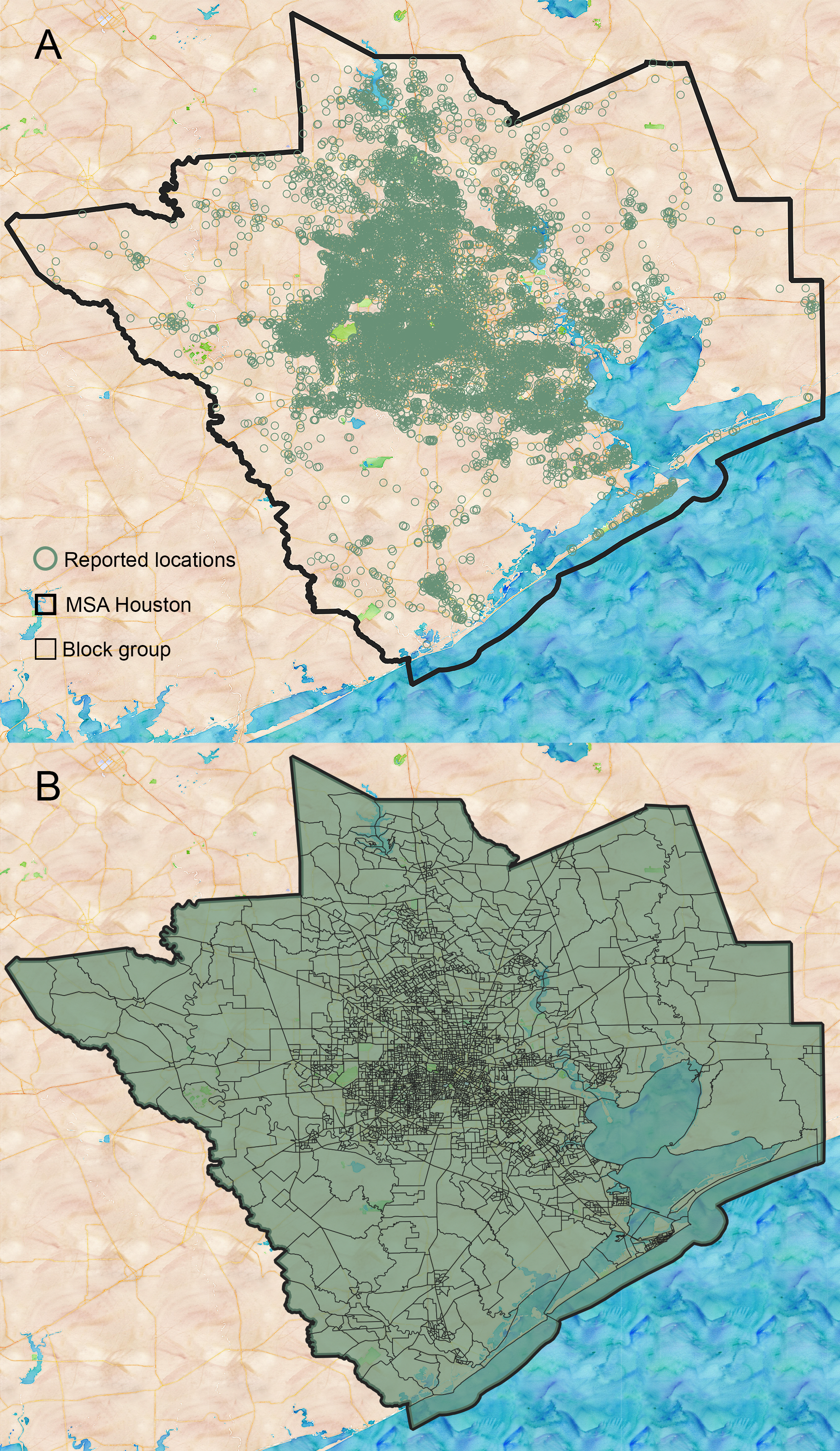} 
    \caption{Data sample and scale. \textbf{\textit{A}}. A sample of 10,000 mobility data in the Metropolitan Statistical Area (MSA) of Houston. Each green circle represents a coordinate reported by a user's smart device. \textbf{\textit{B}}. There are 3,024 block groups in the MSA Houston.}
    \label{fig:fig1}
\end{figure}

The data set used in this study is from August 1 to September 30, 2017, covering the Greater Houston area, i.e., the Houston Metropolitan Statistical Area (MSA) defined by the U.S. Census (Fig. \ref{fig:fig1}). There is about 7 million population in the Houston MSA. Our data includes 5.1 billion data entries from over 2 million devices. All data entries are fully anonymized and collected from opted-in users. Each entry of data contains an anonymized user ID, latitude, longitude, the corresponding time (in seconds), and the precision of the coordinates in meters. The data set has supported interdisciplinary studies on transportation and commuting patterns \cite{wang2019extracting}, urban accessibility \cite{Akhavan2019Accessibility}, mobility reduction and social distancing in COVID-19 \cite{valentino-devries2020location,aleta2020modelling}. Recently, studies have compared results between LBS data sets and other data sets. For example, researchers have analyzed the data set and compared it with transportation survey data \cite{wang2018ondata,wang2019comparative} and mobility patterns observed from Twitter \cite{Phillips2019social}. A sample of 10,000 data points is also shown in Fig. \ref{fig:fig1}A. \par

Besides the mobility data, we also use ACS 2011-2015 Block Group data provided by the U.S. Census. The data set provides socioeconomic information on each block group. There are 3,024 block groups in the Houston MSA and the distribution of the block groups are shown in Fig. \ref{fig:fig1}B.\par

Lastly, we also use geotagged Twitter data from 2013 to 2015 \cite{wang2018urban,Phillips2019social} and Safegraph data set from Feb. 2019~\cite{juhasz2020studying} for comparison purposes. The results from these data sets are mainly reported in the Supplementary Information.\par

\subsection{Home census block groups}
First, we retrieve the stay points of movement trajectories. This step is necessary because similar to other sources of application-based GPS data, our data set, although massive in its volume, can be highly sparse for individuals. It can also contain episodic locations reported when individuals are likely in transition between locations. The stay points are the locations where devices spend sufficient time. We define a stay point using two criteria: (1) duration of at least 15 minutes between location pings and (2) a maximum distance of 50 meters between location pings \cite{zheng2011computing}. \par

Second, we estimate an individual’s home community from the mobility data and link it to their socioeconomic characteristics. We first identify each individual’s home census block groups, using data from Mondays to Thursdays. Since the mobility data set does not keep or share the information, we use hierarchical clustering algorithms to identify each individual’s home neighborhood based on the geographical visits from the phone GPS data (details can be found in \cite{deng2020high}). Although hierarchical clustering is less efficient compared to other algorithms used to identify home neighborhoods in other human mobility studies \cite{jurdak2015understanding,candia2008uncovering,lin_mining_2014,Andrienko2011From,jiang_timegeo_2016}, the algorithm has its advantages on accuracy in identifying clusters and the controllability of cluster sizes \cite{ester1996density,zhu_mapping_2014,ahn_link_2010,rinzivillo_discovering_2012}.  The controllability is especially critical as LBS data often has a relatively high density, i.e., many individuals report their locations every five seconds.  The short intervals between two consecutive points reported by these data can cause the so-called “chaining effect” \cite{hirst_hierarchical_1977,sibson_slink_1973} from which other methods suffer. In such a case, locations could end up in one large cluster even if the start and endpoints are distant from each other. The hierarchical clustering algorithm adopts the complete linkages and thus enforces the maximum diameter of the clusters with a specified maximum distance, and in our case, 50 meters. \par

Individuals' home census block groups might change permanently or temporarily. This is especially true during the strike of Hurricane Harvey as people can evacuate to other places. The change can impact our analysis of mobility data. We develop a dynamic algorithm to find individual's home neighborhoods within a sliding window and it is able to capture the changes. We only keep individuals that have identified home block group for each day of an entire week and the threshold reduces the number of devices to an average of 430,682.9 devices per week. \par

As aforementioned, previous research has found that mobility data can reflect the population distribution to a certain extent. Correlations between the distributions of users identified from various data sets and the one of the general population range from 0.43 to 0.99 depending on the geographical units ~\cite{yabe_effects_2020,wang2019comparative,Sadeghinasr2019Estimating,wang2019extracting}. However, limited research has examined how these correlations can be changed by extreme weather and natural disasters. Based on these findings, we propose Hypotheses 1a and 1b:
\begin{itemize}

    \item \textbf{Hypothesis 1a:} There is a high correlation, i.e., $r>$0.8, between the general distribution of users identified from mobility data and the one of the general population.
    
    \item \textbf{Hypothesis 1b:} There is a high correlation, i.e., $r>$0.8, between the general distribution of users identified from mobility data and the one of the general population during the occurrence of Hurricane Harvey.
\end{itemize}

\subsection{Calculating Justice Measures for Hypothesis Tests} 
Even if we find support for our first set of hypotheses, i.e., the general distributions can be highly correlated, the representativeness among different types of neighborhoods can vary. Previous research has argued that representativeness from certain social media is skewed \cite{mislove2011understanding,samuels2020silence,fan2020spatial,Zou2019Social}. It has also been reported that minorities are often not represented in crowdsourced data sets \cite{Wang2016Process}. Since disadvantaged communities can have a higher level of vulnerability and thus suffer from more damages during natural disasters, the representativeness can change during the occurrence of a natural disaster. We categorize the neighborhoods in the greater Houston area in two ways. The first way is based on race. A majority white neighborhood has more than 50\% of its population as non-Hispanic whites. The same simple majority threshold applies to majority black and Hispanic neighborhoods. The second way is by poverty level. If a neighborhood has more than 30\% of its population living below the poverty line, it is classified as a poor neighborhood; otherwise, a nonpoor one. \par 

We measure representativeness, our first justice parameter, using the ratio $p_i$ shown in the following Eq. (\ref{representativeness})
\begin{equation}
    p_i=\left(\frac{x_i}{N_i}\right)
\label{representativeness}
\end{equation}

\noindent where $x_i$ is the identified individuals in block group $i$ and $N_i$ is the total population in the same block group reported by the U.S. Census. \par

Eq. (\ref{representativeness}) can be modified for different types of neighborhoods. Eq. (\ref{representativeness_w}) measures the representativeness in a majority white neighborhood:

\begin{equation}
    p_{i,w}=\left(\frac{x_{i,w}}{N_{i,w}}\right)
\label{representativeness_w}
\end{equation}

The measure allows our test on the second set of hypotheses:\par

\begin{itemize}

\item \textbf{Hypothesis 2a:} $p_{i,w}>p_{i,b}$.

\item \textbf{Hypothesis 2b:} $p_{i,w}>p_{i,h}$.

\item \textbf{Hypothesis 2c:} $p_{i,w}>p_{i,b}$ during the occurrence of Hurricane Harvey.

\item \textbf{Hypothesis 2d:} $p_{i,w}>p_{i,h}$ during the occurrence of Hurricane Harvey.

\item \textbf{Hypothesis 2e:} $p_{i,np}>p_{i,p}$.

\item \textbf{Hypothesis 2f:} $p_{i,np}>p_{i,p}$ during the occurrence of Hurricane Harvey.
\end{itemize}

The injustice can go beyond the representativeness. As mentioned above, two factors can also impact the mobility data justice: quantity and precision. An understanding of both of them is important as mobility data has been used to predict travels and their associated human activities \cite{song2010limits,cuttone2018,wang2017predicting,liu2018urban,feng2018deepmove}. These studies developed mathematical models, machine learning and deep learning approaches, which depend on the quantity and precision of input. Low quantity and precision of mobility data can cause biased engineering solutions and negligence on vulnerable populations' needs in emergencies. \par

Instead of focusing on the number of entries from the raw data, we focus on stay points to develop our quantity measures. The rationale is that stay points are where people spend time, representing their visitations in the urban area. Using the stay points we estimated, we calculate two measures: (1) $q_{h,i}$ which is the average number of hours out of the 24 hours that individuals from block group $i$ has stay points, and (2) $q_{sp,i}$ which is the average number of stay points of the individuals from block group $i$. \par 

Using the two measures, we test the following two sets of Hypotheses:

\begin{itemize}
    \item \textbf{Hypothesis 3a:} $q_{h,w} > q_{h,b}$ .

    \item \textbf{Hypothesis 3b:} $q_{h,w} > q_{h,h}$.
    
    \item \textbf{Hypothesis 3c:} $q_{h,w} > q_{h,b}$ during Hurricane Harvey.
    
    \item \textbf{Hypothesis 3d:} $q_{h,w} > q_{h,h}$ during Hurricane Harvey.
    
    \item \textbf{Hypothesis 3e:} $q_{h,np} > q_{h,p}$.
    
    \item \textbf{Hypothesis 3f:} $q_{h,np} > q_{h,p}$ during Hurricane Harvey.
\end{itemize}

\begin{itemize}
    \item \textbf{Hypothesis 4a:} $q_{sp,w} > q_{sp,b}$ .

    \item \textbf{Hypothesis 4b:} $q_{sp,w} > q_{sp,h}$.
    
    \item \textbf{Hypothesis 4c:} $q_{sp,w} > q_{sp,b}$ during Hurricane Harvey.
    
    \item \textbf{Hypothesis 4d:} $q_{sp,w} > q_{sp,h}$ during Hurricane Harvey.
    
    \item \textbf{Hypothesis 4e:} $q_{sp,np} > q_{sp,p}$.
    
    \item \textbf{Hypothesis 4f:} $q_{sp,np} > q_{sp,p}$ during Hurricane Harvey.
\end{itemize}

Lastly, we develop a precision measure. We use the median of the precision values $\hat{\mu_i}$ from the individuals of a block group. As before, we calculated the measure for majority white, black, Hispanic, poor, and nonpoor neighborhoods. We use the precision measure to test the following hypotheses: \par 

\begin{itemize}
    \item \textbf{Hypothesis 5a:} $\hat{\mu_w} > \hat{\mu_b}$.

    \item \textbf{Hypothesis 5b:} $\hat{\mu_w} > \hat{\mu_h}$.
    
    \item \textbf{Hypothesis 5c:} $\hat{\mu_w} > \hat{\mu_b}$ during Hurricane Harvey.
    
    \item \textbf{Hypothesis 5d:} $\hat{\mu_w} > \hat{\mu_h}$ during Hurricane Harvey.
    
    \item \textbf{Hypothesis 5e:} $\hat{\mu_{np}} > \hat{\mu_p}$.
    
    \item \textbf{Hypothesis 5f:} $\hat{\mu_{np}} > \hat{\mu_p}$ during Hurricane Harvey.
\end{itemize}

\section{RESULTS}
\begin{figure}[htbp]
    \centering
    \includegraphics[page=1,width=\textwidth]{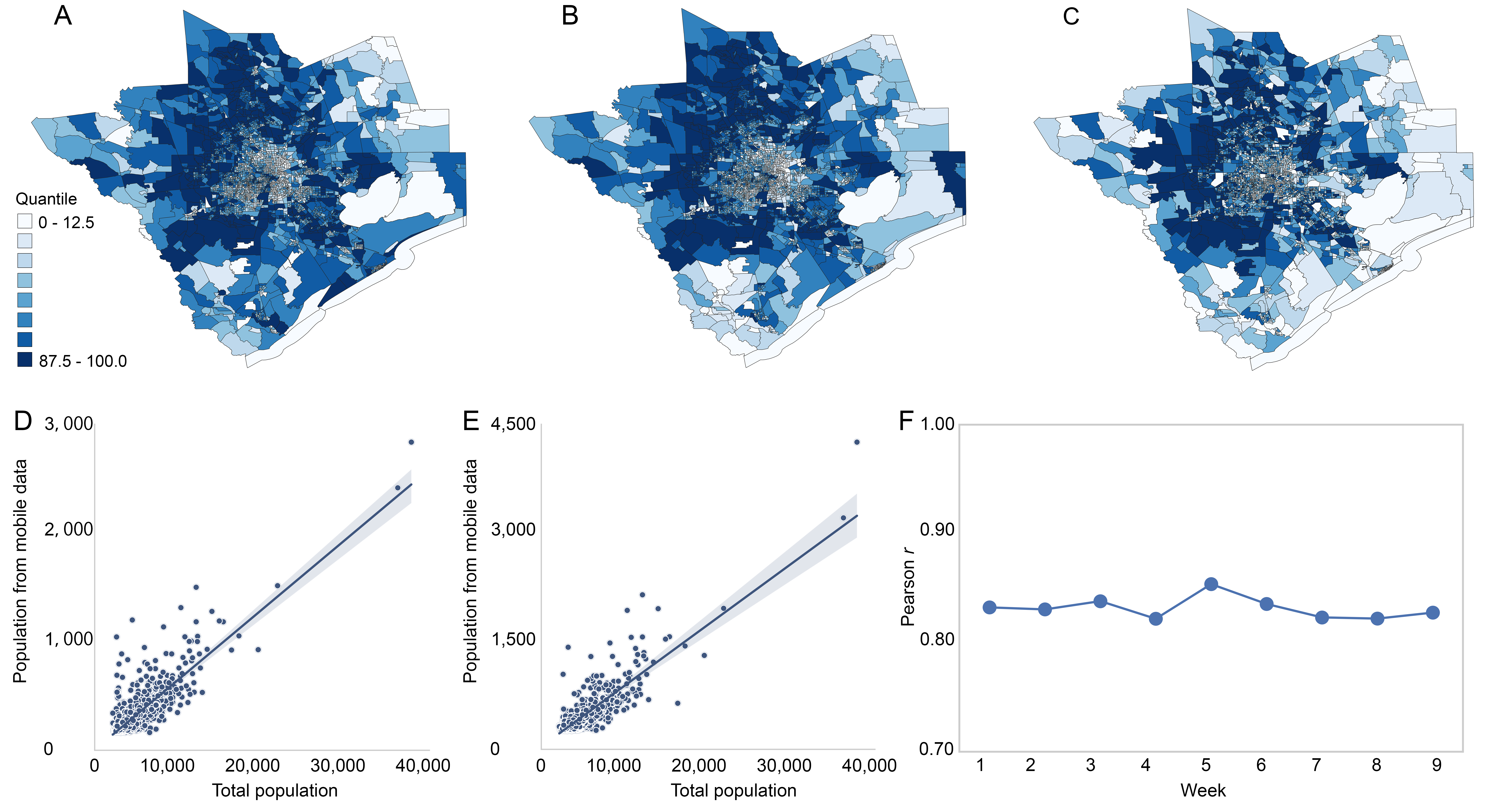} 
    \caption{The correlation between users from mobility data and the general population. \textbf{\textit{A}}. The distribution of the numbers of users from the mobility data in the block groups during the week of July 31 to August 3. The colors indicate the quantiles of the numbers of devices, with the lightest blue represents 0 to 12.5\% quantile and the darkest represents 87.5\% to 100\% quantile.  \textbf{\textit{B}}. The distribution of the same number during the week of August 28 to August 31. \textbf{\textit{C}}. The distribution of the population reported by the U.S. Census. \textbf{\textit{D}}. The correlation between \textbf{\textit{A}} and \textbf{\textit{C}}. The Pearson \textit{r} is 0.83***. \textbf{\textit{E}}. The correlation between \textbf{\textit{B}} and \textbf{\textit{C}}. The Pearson \textit{r} is 0.85***. \textbf{\textit{F}}. The changes of \textit{r} during the nine weeks before, during, and after the strike of Hurricane Harvey.}
    \label{fig:fig2}
\end{figure}

Fig. \ref{fig:fig2}A shows the distributions of the numbers of identified devices in the first week, and Fig. \ref{fig:fig2}B the number during week 5 when Hurricane Harvey made landfall. In comparison, the distribution of the general population can be found in Fig. \ref{fig:fig2}C. The correlations of the identified individuals and general populations are $r$ = 0.83*** in Week 1 (Fig. \ref{fig:fig2}D) and $r$ = 0.82*** in Week 5 Fig. (\ref{fig:fig2}E) respectively. The values of Pearson $r$ in all the weeks are shown in Fig. \ref{fig:fig2}F. \par

\begin{figure}[htbp]
    \centering
    \includegraphics[page=1,width=0.8\textwidth]{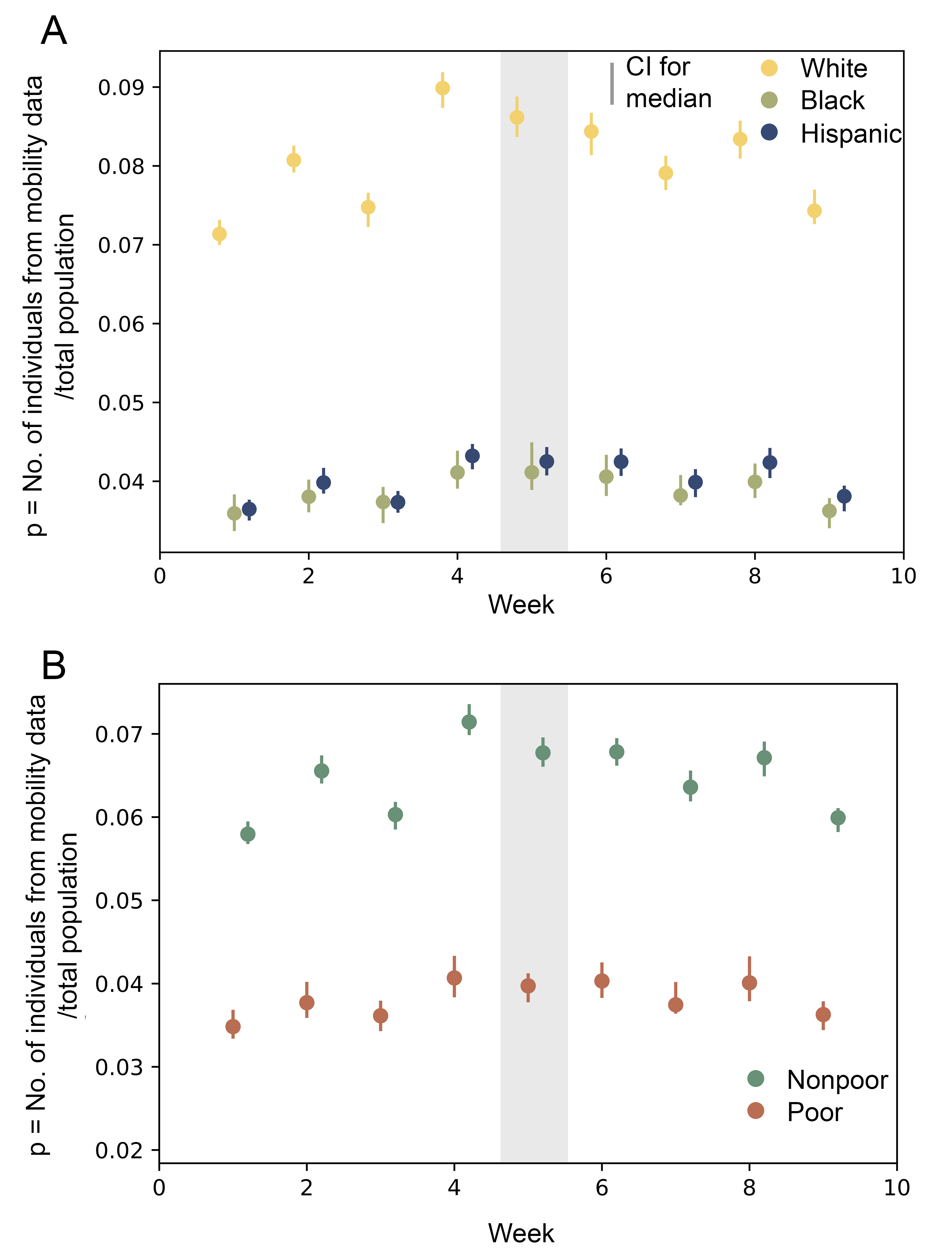} 
    \caption{The representativeness of the users from the mobility data. \textbf{\textit{A}}. The representativeness for dominantly white (yellow), black (olive), and Hispanic (blue) neighborhoods during the nine weeks. \textbf{\textit{B}}. The representativeness for dominantly nonpoor (green) and poor (red) neighborhoods during the nine weeks. The dots show the medians and the bars show the 95\% confidence intervals around the medians.}
    \label{fig:fig3}
\end{figure}

While $r$ remains high through the progress of the natural disaster, the values of $p$ are different across different types of neighborhoods. Figure \ref{fig:fig3}A shows the changes of $p$ across the 9 Weeks. The dots represent the median values and the bars represent the 95\% confidence intervals (CIs) around median values (see \cite{bland2015introduction} for the method of calculation). In Week 1, the median $p_w$ was 0.071 for dominantly white neighborhoods. The values were 0.036 and 0.036 for dominantly black ($p_b$) and Hispanic ($p_h$) neighborhoods respectively. The medians values for nonpoor and poor neighborhoods were $p_p$=0.035 and $p_{np}$=0.058 respectively (Figure \ref{fig:fig3}B) in Week 1. In Week 5, the values increased and results were $p_w$=0.085, $p_b$=0.042, $p_h$=0.043, $p_p$=0.040, $p_{np}$=0.068. The CIs of white neighborhoods have no overlaps with the ones of black and Hispanic neighborhoods, and the Mood's median tests report high significance values (the p-values can be found in Table A1). \par

\begin{figure}[htbp]
    \centering
    \includegraphics[page=1,width=\textwidth]{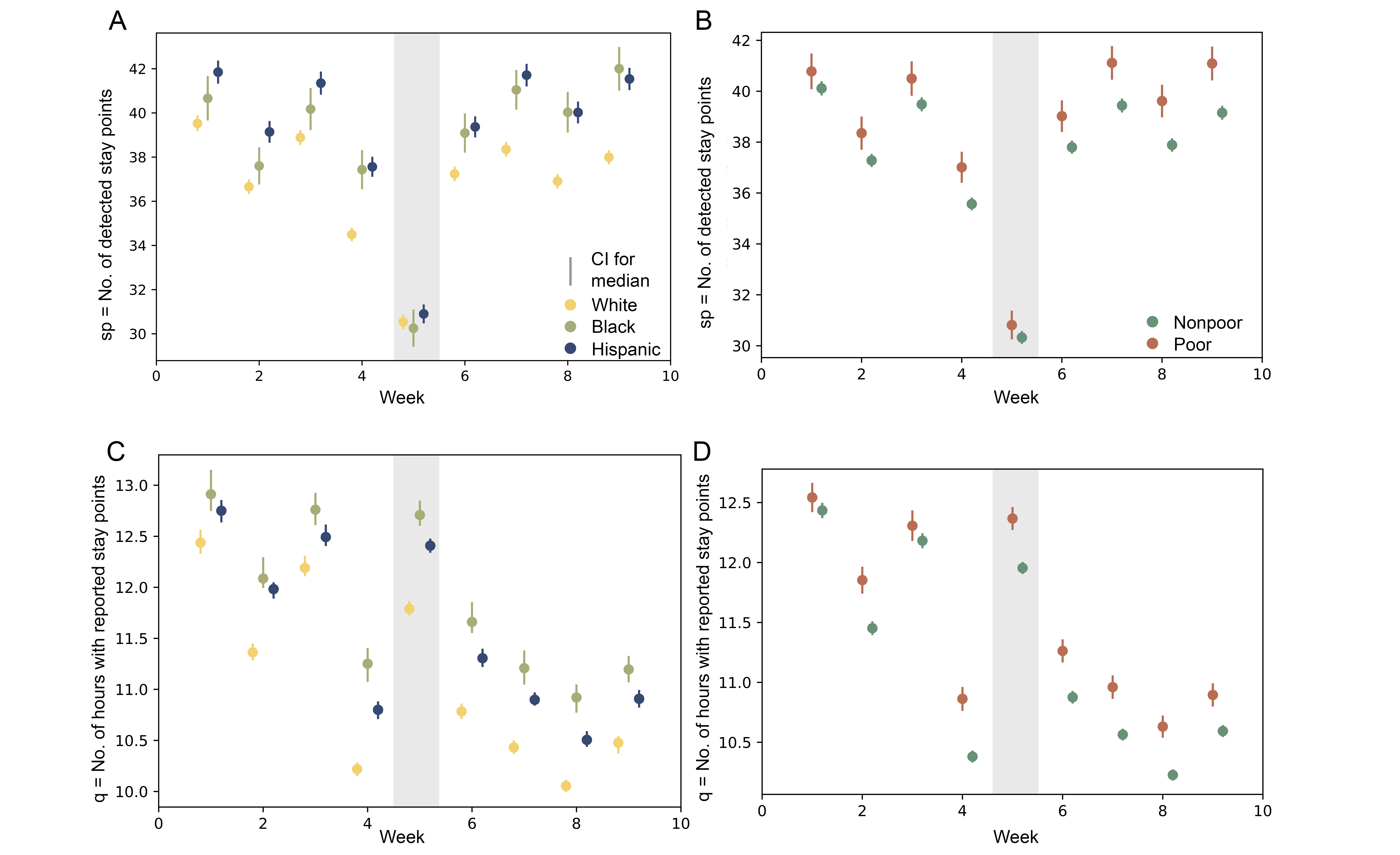} 
    \caption{The quantity of mobility data. \textbf{\textit{A}}. The number of hours ($q$) reported from the stay points for dominantly white (yellow), black (olive), and Hispanic (blue) neighborhoods during the nine weeks. \textbf{\textit{B}}. The number of hours ($q$) reported from the stay points for dominantly nonpoor (green) and poor (red) neighborhoods during the nine weeks. \textbf{\textit{C}}. The numbers of stay points ($sp$) for dominantly white (yellow), black (olive), and Hispanic (blue) neighborhoods during the nine weeks. \textbf{\textit{D}} shows the number of stay points ($sp$) for dominantly nonpoor (green) and poor (red) neighborhoods during the nine weeks. The dots bars show the medians in all four panels, and the bars show the 95\% confidence intervals around the medians.}
    \label{fig:fig4}
\end{figure}

Fig. \ref{fig:fig4}A and B show the numbers of hours with reported data $q$ during the occurrence of Hurricane Harvey. The values of $q$ ranged from 10 to 14 hours and remained consistent throughout the nine weeks. Fig. \ref{fig:fig4}C and D show the change in the numbers of stay points $q_sp$. There were no significant differences among the neighborhoods of varying sociodemographic groups even though the numbers of $q_sp$ reduced significantly during the week when Hurricane Harvey struck. \par

\begin{figure}[htbp]
    \centering
    \includegraphics[page=1,width=0.8\textwidth]{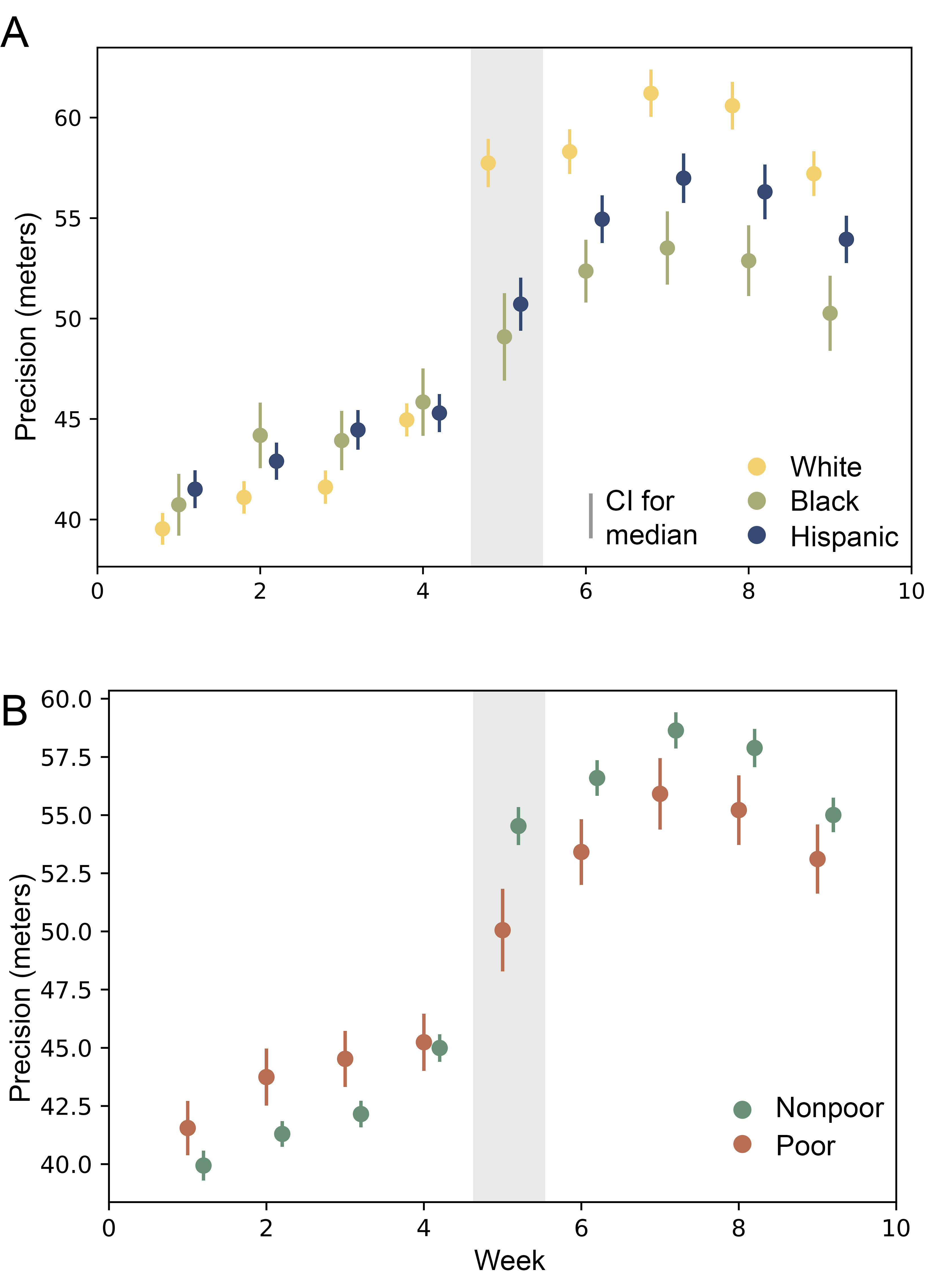} 
    \caption{The precision of mobility data. \textbf{\textit{A}}. The data precision for dominantly white (yellow), black (olive), and Hispanic (blue) neighborhoods during the nine weeks. \textbf{\textit{B}}. The data precision for dominantly nonpoor (green) and poor (red) neighborhoods during the nine weeks. The dots show the medians and the bars show the 95\% confidence intervals around the medians. }
    \label{fig:fig5}
\end{figure}

The precision results are shown in Fig. \ref{fig:fig5}. The median precision from white neighborhoods remained around 20 meters before the land of Hurricane Harvey and then increased to 27.91 meters in Week 5 and 34.04 meters in Week 6. After that, the precision starts to recover. We observe similar median values, as well as changes, in black and Hispanic neighborhoods (Fig. \ref{fig:fig5}A) as well as in poor and nonpoor neighborhoods (Fig. \ref{fig:fig5}B). 

\section{DISCUSSIONS}

Our analytical results confirm that mobility data injustice vary in the three parameters tested: representativeness, quantity, and quality. In general, the differences are more substantial in representativeness. We also observe specific changes in the quantity and quality of mobility data caused by Hurricane Harvey with important implications. We discuss them in the order of our hypotheses presented in Section 3. \par

In the first set of hypotheses, we expect a high correlation between the general distribution of users identified from mobility data and the one of the general population. We assume that the hypothesis is true for both before and during Hurricane Harvey. Our analysis of the empirical data shows that the values of Pearson's $r>0.8$ across different weeks with $p<0.001$. Thus, we reject the null hypotheses and find support for both Hypotheses 1a and 1b. It is also worth pointing out that the high correlation is obtained on the block group level, which is the smallest geographical unit with robust estimates on sociodemographic composition from the U.S. Census. We also tested on the tract level, one level above block groups. The 3,021 block groups in Houston MSA comprise 1,070 tracts. The correlation remains high as $r$=0.82 in week 1 and 0.85 in week 5. Also, for the entire 9-week period when considering all home block groups, the estimated $r=0.94$ on the block group level and $r=0.95$ on the Tract level. The findings align with previous studies ~\cite{yabe_effects_2020,wang2019comparative,Sadeghinasr2019Estimating} and support the values of using similar data sets to study mobility patterns on aggregated levels. \par

In our second set of hypotheses, we assumed the representativeness is higher in neighborhoods with dominantly white population than the values in the neighborhoods with dominantly black and Hispanic populations. The estimated $p$ confirms our hypotheses 2a and 2b. As shown in Figure \ref{fig:fig3}A, there is a higher proportion of individuals identified that reside in majority-white block groups. The values of $p_w$ range from 7.1\% to 9.0\%. On the other hand, $p_b$ ranges from 3.6\% to 4.1\% in majority-black block groups and $p_h$ ranges from 3.6\% to 4.3\% in majority Hispanic block groups. Similarly, we find support for Hypotheses 2e. The values of $p_{np}$ range from 5.8\% to 7.1\% while the ones of $p_p$ range from 3.5\% to 4.1\% (Fig. \ref{fig:fig3}B).\par

Although an extensive comparison is beyond the scope of this study, we compared the results with data from geotagged Twitter data from 2013 to 2015 used in \cite{wang2018urban,Phillips2019social} and Safegraph data set from Feb. 2019 \cite{juhasz2020studying}. The Twitter data includes 12,984 users, and the methods for finding home locations and residential neighborhoods are discussed in detail in \cite{wang2018urban}. The small number of users leads to low representativeness $p$ overall, ranging from 0.2\% to 0.8\%, for the geotagged Twitter data (see Fig. A1). Probably owing to such overall low representativeness, the disparity is not clear in Twitter data even though it covers a longer period. On the other hand, Safegraph identifies "home" locations for 588,563 users using the "common" nighttime location, although the exact method is not explained in detail. We find that the pattern observed from Safegraph data sets aligns with what we observed in our mobility data set (Fig. A2). The differences are still statistically significant yet less substantial. The alignment between the two results highlights the potential justice issues in crowdsourced data sets.\par

The disparity persisted during the strike of Hurricane Harvey. In Week 5 (Fig. \ref{fig:fig3}A and B shadowed area), $p_w$=8.6\%, $p_b$=4.1\%, $p_h$=4.3\%,$p_{np}$=6.8\% and $p_{p}$=4.0\%. Therefore, Hypothesis 2c, 2d, and 2f are confirmed. It is worth noting that the representativeness tends to increase in Week 4 and Week 5, right before and during Hurricane Harvey's landfall. The increases are universal across all types of neighborhoods and can be attributed to decreased human mobility when Houstonians were preparing and facing the natural disaster. As pointed out by previous research, individuals tend to limit their mobility to short-distance travel and reduce long-distance travel ~\cite{wang_patterns_2016,wang_quantifying_2014}, which is likely to generate more data points in one's residential neighborhoods. Thus, our algorithm detects more homes within the residential block groups. \par

Our third and fourth sets of hypotheses build upon the assumption that the quantity of data from advantaged neighborhoods is higher than the one from disadvantaged neighborhoods. We calculated two measures, active hours and number of stay points, to test our hypotheses. The CIs (Fig. \ref{fig:fig4}) and the results from the Mood's median tests (see Table A1) provide no support for these hypotheses. Two findings are worth highlighting. Firstly, contradicting what we hypothesized, numbers of stay points $q_{sp}$ and active hours $q_h$ from black neighborhoods and Hispanic neighborhoods are statistically significantly higher than those from white neighborhoods in most of the weeks before Hurricane Harvey. We do observe these differences are not substantial. On average, $q_{h,w}$ is 11.55 hours, and $q_{h,b}$ is 12.25 hours in the first four weeks, an increase of 0.72 hour (6.1\%). Similarly, $q_{h,h}$ is 12.01 hours, an increase of 0.43 (3.9\%) hour comparing to $q_{h,w}$. For the stay points, $q_{sp,b}$ (38.97) and $q_{sp,h}$ (39.97) are 4.2\% and 6.9\% higher than $q_{sp,w}$ (37.39). The differences in Week 6 to Week 9 are on a par with the ones in the first four weeks. The differences of $q_{sp}$ and $q_h$ are also statistically significant in most of the weeks and yet not substantial. The results align with previous findings. For example, Wang et al. ~\cite{wang2018urban} reported that populations from disadvantaged neighborhoods tend to visit a similar number of neighborhoods when comparing to people from "main-stream" neighborhoods. \par

Another finding from this test is that the numbers of distinct visits, i.e., stay points, $q_{sp}$ experience substantial changes during the strike of Hurricane Harvey. When comparing $q_{sp}$ in Week 5 with the average values in other weeks, we observe a significant drop, as high as 22.7\% during the occurrence of Hurricane Harvey (shaded regions in Fig. \ref{fig:fig4}C and D). The Mood's median tests also show that the differences between neighborhoods are not as significant as before the landfall. The results indicate that residents from all neighborhoods experienced disruptions and are forced to reduce their mobility significantly. \par

Our last set of hypotheses are developed to test the data precision $\hat{\mu_i}$ of different neighborhoods. We find that $\hat{\mu_w}$ is slightly smaller than $\hat{\mu_b}$ and $\hat{\mu_h}$ from Week 1 to Week 4 (Fig. \ref{fig:fig5}A), and the differences are only statistically significant in Week 2 and 3 (Table A1). The precision values started to drop for all neighborhoods as $\hat{\mu}$ began to increase at Week 5. Surprisingly, $\hat{\mu_w}$ surpassed both $\hat{\mu_b}$ and $\hat{\mu_h}$ and was statistically higher in Week 5 to 9. Therefore, Hypothesis 5a to 5d are not supported. We observed a similar trend when comparing poor and nonpoor neighborhoods. Thus, we cannot reject the null hypotheses and find no support for Hypotheses 5e and 5f. \par

Despite finding no support for the fifth set of hypotheses, we observe an important phenomenon: the mobility data's precision decreased significantly during and after Hurricane Harvey. It started to drop in Week 5 and reached the lowest point in Week 7, losing 54.8\% precision. The precision levels then gradually returned but not to the same level before the natural disaster by Week 9. The findings have important implications for disaster response and relief. First responders might rely on LBS data to locate individuals, identify people in need, and allocate resources. The decrease of data precision can impact these efforts and add difficulties in responding to emergencies and providing precise and accurate post-disaster recovery needs.  \par 

\section{LIMITATION}

We focused our analysis on arguably one of the most comprehensive data sets used to study commuting patterns, urban accessibility, mobility, and social distancing in COVID-19 (see Section 3.2). The popularity of the data set warrants the in-depth analysis presented in this study. Although we found the results align with patterns observed in another data set, i.e., the Safegraph data, future studies should examine more mobility data sets. Future studies can also benefit from reviewing more natural disasters and man-made extreme events. The patterns observed in this study can be altered in other types of events. \par

\section{CONCLUSION}
Big data not only deserves a big audience \cite{huberman2012big} but also needs to include equal and representative contributors. Data crowdsourced from smart devices could help us better respond to emergencies \textit{only if} it gives voices to minorities, disadvantaged, and vulnerable populations. This study took one of the first steps to examine the mobility data justice using 2017 Hurricane Harvey as a case study. Our findings show that a persistent disparity of representativeness was observed before, during, and after the hurricane's landfall. The representativeness was significantly and substantially higher in majority-white and non-poor neighborhoods when compared to majority-black, -Hispanic, and poor neighborhoods. Additionally, we observed significant drops of data precision across different types of neighborhoods, adding uncertainty to locate people and understand their movements during extreme weather events. The findings indicate that research and applications based on mobility data must consider and control potential biases and justice issues. \par

Despite revealing potential justice issues in mobility data, this study is by no means an attack on or denial of the value of crowdsourced data sets. As we discussed in Section 2, mobility data sets have supported answering many critical research questions. As Femke \cite{Femke2020Humanitarian} pointed out, "data justice is a complex and multidimensional problem marked by multiple interlinking elements." Data injustice is rooted in social injustice and demands systematic solutions. Future studies should focus on developing mitigation strategies to address injustice issues. Also, a data justice lens, such as the one proposed in \cite{Femke2020Humanitarian}, could provide practical tools to guide the collection and analysis of data during disasters. \par

\section*{ACKNOWLEDGMENT}
Our data usage agreement does not allow us to make public or otherwise share the anonymized mobile phone data collected from opted-in users in this study. Researchers interested in aggregated data and/or summary statistics, where permitted under said agreement, should contact the corresponding author. The code used to generate the results of this paper is available from the corresponding authors upon request.

\bibliographystyle{elsarticle-num} 
\bibliography{scibib.bib}

\end{document}